\documentstyle[twoside,fleqn,espcrc2,psfig]{article}

\newcommand{\AmS}{{\protect\the\textfont2
  A\kern-.1667em\lower.5ex\hbox{M}\kern-.125emS}}

\newcommand{\pom}{I\!\!P}

\newcommand{\ggx}{\gamma\gamma }
\newcommand{\gstar}{\gamma^*\gamma^* }
\newcommand{\ee}{e^+e^-}
\newcommand{\Wgg}{W_{\gamma\gamma}}

\title{Total $\ggx$ 
and $\gstar$ 
Cross Sections Measured at  LEP}

\author{Albert De Roeck\address[CERN]{
CERN, 1211 Geneva 23, Switzerland}
}
       
\begin{document}

\begin{abstract}
Recent results on total cross-section measurements in $\ggx$ and $\gstar$
collisions at LEP are reported. Phenomenological fits to the data 
are presented.
\vspace{1pc}
\end{abstract}

\maketitle

\section{Introduction}

At LEP2 photon-photon collisions constitute a large part
of the inclusive cross-section.
Quasi real photons are emitted according to a
Weizs\"acker-Williams energy spectrum
by the lepton beams. Two photon events can  be tagged 
by one or both of the scattered electrons, or anti-tagged, in 
which case the 
electrons remain in the beampipe. Tagging detectors measure electrons 
typically down to 30 mrad. In this paper we report results on 
anti-tagged (i.e.~almost real photon $\ggx$) and double tagged
(i.e.~$\gstar$) total cross-section measurements.
The analyses are based on approximately 390 pb$^{-1}$, from
$e^+e^-$ data taken at $\sqrt{s_{ee}} = 189-202$ GeV in '98 and '99.

\section{Total $\gamma\gamma$ Cross-Section}
Photons at high energies can fluctuate in two-fermion pairs or
vector mesons with the same $J^{PC}$ quantum numbers as for the photon.
Hence a photon can behave like a hadron, with an additional
pointlike component.
Total inclusive hadron-hadron cross-sections are known to rise
  gently with the centre-of-mass (CMS) energy.
An outstanding question is whether $\sigma_{\gamma\gamma}$ 
rises  faster
than $\sigma_{pp}$, expected from  
the additional  pointlike component in the photon structure.


LEP allows to study $\gamma\gamma$ interaction cross-section as function
of the CMS energy $W_{\gamma\gamma}$.
 L3~\cite{L3sig} and OPAL~\cite{OPALsig} have measured
$\sigma_{\gamma\gamma}$ in the region $5 < \Wgg < 145$ GeV
and $10 < \Wgg < 110$  GeV respectively.
The challenge of this measurement is the reconstruction of
$W_{\gamma\gamma}$ from the visible hadronic
final state. The result depends on the Monte Carlo model 
used to correct the data: it affects the absolute normalization by
approximately 20\%, as derived from using two different models, 
PYTHIA~\cite{pythia} and PHOJET~\cite{phojet}.
An important uncertainty is the diffractive contribution to the 
cross-section. Such events to a large extent escape detection;
depending on the model only 6\% to 20\% of the diffractive 
events are selected by the standard  analysis cuts.

\begin{figure}[htbp]
\begin{center}
\psfig{figure=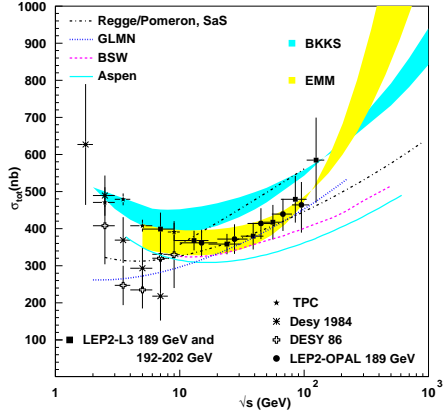,bbllx=-30pt,bblly=190pt,bburx=420pt,bbury=620pt,width=5cm}
\caption{ Data on $\sigma_{\gamma\gamma}$, versus $W_{\gamma\gamma}$ 
(denoted as $\sqrt{s}$)
compared with models (see text).}
\label{fig1}
\end{center}
\end{figure}

In 
Fig.~\ref{fig1} data on  $\sigma_{\gamma\gamma}$ from the LEP and 
pre-LEP experiments are shown, compared to various theoretical models
as reviewed in~\cite{pancheri}. The new L3 data are still preliminary and
are a combination of data taken at $\sqrt{s_{ee}} = 189$ and $192-202$ 
GeV~\cite{L3paper}.
Contrary to previously published data~\cite{L3sig}, these data are 
determined using both PHOJET and PYTHIA, and are therefore
above published values.
The cross-section clearly rises
with increasing $W_{\gamma\gamma}$. 
The OPAL and L3 data are consistent with each other.
This is less so for the pre-LEP data~\cite{soldnerDIS}, 
of which only a selection is
shown in the figure.

All models predict a rise with the collision 
energy, but the strength of the rise
differs and 
the predictions at high  energy show dramatic
differences. 
In {\it proton-like-models}
(dash-dotted~\cite{SAS,us1,bib-DL}, dashed~\cite{ttwu},
dotted~\cite{GLMN} and 
solid~\cite{aspen} curves), the curvature 
follows  closely that of proton-proton cross-section, in {\it QCD based} 
models (upper~\cite{BKKS} and lower~\cite{pancheri} bands), the rise is 
obtained using the pQCD jet cross-section. 

\begin{figure}[htbp]
\begin{center}
\psfig{figure=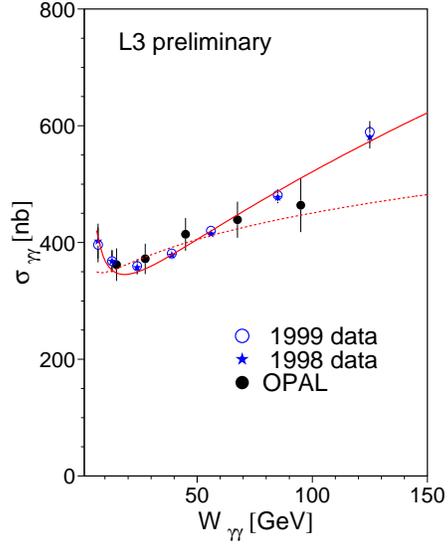,bbllx=-20pt,bblly=30pt,bburx=460pt,bbury=650pt,width=5cm}
\caption{Data on $\sigma_{\gamma\gamma}$ from OPAL and L3, compared with
fits through the data (see text).}
\label{fig3}
\end{center}
\end{figure}

A large contribution to the errors of the cross-section is due to
the uncertainty of the model dependent correction
factors, which are  strongly  point-to-point correlated and partially 
hide the significance of the 
 rise of the cross-section. Therefore
Fig.~\ref{fig3} shows the L3 data without the model dependent 
errors.
The size of the rise is quantified 
by a Regge inspired fit, $X_1\cdot s^{\epsilon_1} +
Y_1\cdot s^{-\eta_1}$, i.e. the sum of a pomeron and a Reggeon 
 term. 
The exponents $\epsilon_1= \alpha_{\pom}-1$ and 
$\eta_1=\alpha_{R}-1$ are usually assumed to be universal,
whereas the coefficients are process dependent.
For $\eta_1$, which is determined by low energy hadronic data, 
the value of 0.34 is taken~\cite{bib-pdg}. $\alpha_{\pom}$ and $\alpha_{R}$
are the pomeron and Reggeon intercept, respectively.

For its fits,
OPAL fixed the coefficient of the Reggeon term, while L3 fitted the
magnitude of $Y_1$. The L3 fit yields for the pomeron term:
$\epsilon_1 = 0.263\pm0.014$~\cite{L3paper}
while OPAL published $\epsilon_1 = 
0.101^{+0.025}_{-0.020}$~\cite{OPALsig}. The curves show the 
L3 fits with free (solid line) and fixed (dashed line) value
of $\epsilon_1, (\epsilon_1=0.095)$.
Correlations between the points are taken into account in these fits.
L3 observed that the value of $\epsilon_1$ does not depend
significantly  on the 
model used for correcting the data.
The slope found by OPAL is considerably smaller than the L3 one,
despite agreement between the data points.
This can be traced back to the assumption on the Regge 
coefficient, $Y_1$, which 
is fixed by OPAL due to absence of its own measurements below 
$\Wgg = 10$ GeV to a value of 320 nb. L3 finds $Y_1 \sim 1000$ nb
from its own fits.
Hence we made fits  using the published OPAL and preliminary new 
L3 data points, ignoring  correlations between the points and
ignoring the uncertainty due to the model dependence.
Refitting the OPAL data but
with
$Y_1$ as determined by L3, one finds
$\epsilon_1 = 0.205\pm 0.042$ $ (\chi^2/NDF = 1/3) $. 
Refitting all the data, leaving $Y_1$ free, gives
$\epsilon_1 = 0.238\pm 0.029$ $ (\chi^2/NDF = 2.7/9)$.
Is this rise driven by the point at largest $\Wgg$ from L3?
Removing this point from the fit gives $\epsilon_1 = 0.223\pm 0.033 $, hence
no significant change of the exponent.

The corresponding value for $\epsilon_1$ in hadronic collisions is 
in the range $0.08-0.095$, hence
the $\gamma\gamma$ cross-section  
appears to rise significantly faster  than hadron cross-sections.

Recently, Donnachie-Landshoff~\cite{bib-DL2} proposed 
a model which includes two pomeron terms in 
an attempt 
to save the universality of the soft pomeron.
The total cross-section is then assumed to be described by 
\begin{equation}
\sigma=X_1 s^{\epsilon_1}+X_2 s^{\epsilon_2}+Y_1 s^{-\eta_1},
\label{eq-tot1}
\end{equation}
The second pomeron term was added mainly to  describe the $\gamma^*p$ data.
From fits to $pp$ and $\gamma^{(*)} p$ data 
the exponents
$\epsilon_1 = 0.0808$ $\epsilon_2 = 0.418$ and $\eta_1 = 0.34 $   
are extracted.
 An intriguing question is whether 
the $X_2$ term is present in the $\gamma\gamma$ data.
A fit, as described above, to the 
preliminary L3 data with $\Wgg > 20 $ GeV, and with two pomerons only,
gives $X_2= 2.5\pm 0.6$ nb, with $\chi^2/NDF$
= 4.5/3.
Fitting all  OPAL and L3 data to the full expression of
eq.(~\ref{eq-tot1}),
keeping the $\epsilon_1,\epsilon_2$
and $\eta_1$ fixed, gives $X_2= 5.0\pm 0.9$ nb, with $\chi^2/NDF
=2.3/9$.
Hence, within this model, and within the restrictions 
of the fits made, the extra hard pomeron term appears 
to be  required 
in  $\gamma\gamma$ data at the $4\sigma$ level or higher.

While the cross-section in $\gamma\gamma$ appears to be rising faster than 
in $pp$,  the $\gamma p$ cross section can be described  by 
a single pomeron term with exponent  0.0808. 
It is therefore important that the 
rise in the $\gamma\gamma$ data gets 
established more thoroughly. This can be accomplished with more and improved
cross-section measurements in the high $\Wgg$ range 
AND also with better measurements
at low $\Wgg$ from LEP or elsewhere. 
The latter is important because the fit result depends significantly on the 
knowledge of the Reggeon term.
An experiment at VEPP in 
Novisibirsk has been scheduled  to measure
$\sigma_{\gamma\gamma}$ at low $\Wgg$.

\section{Total $\gamma^*\gamma^*$ Cross-Section}

The study of  virtual photon-photon scattering has recently been discussed
as a definite probe of the hard (BFKL) pomeron~\cite{BFKL}.
The BFKL  evolution equation describes scattering processes in QCD in the
limit of large energies and fixed, but sufficiently large momentum transfers.
The BFKL pomeron
has been intensively investigated in the context
of small--$x$ deep inelastic electron-proton scattering at HERA.
However, despite clear hints~\cite{deroeck},
the presence of so called $\ln 1/x$ terms in
deep inelastic scattering at HERA
has not been unambiguously established  yet.


\begin{figure}[htbp]
\begin{center}
\psfig{figure=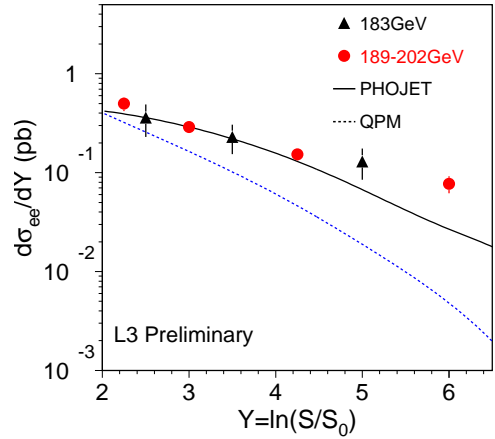,bbllx=-30pt,bblly=20pt,bburx=450pt,bbury=430pt,width=5.5cm}
\caption{The $ee\rightarrow eeX$ cross-section for double tagged events,
measured by L3,
compared with model predictions (see text).}
\label{fig5}
\end{center}
\end{figure}

\begin{figure}[htbp]
\begin{center}
\psfig{figure=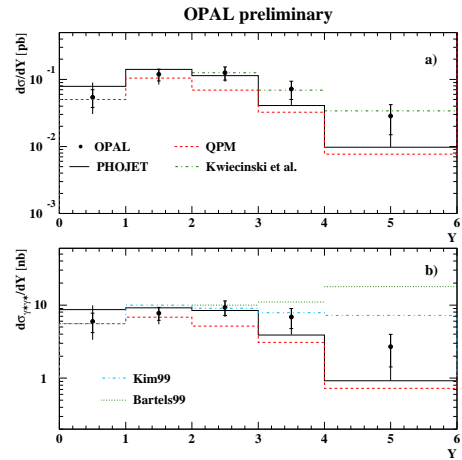,bbllx=-40pt,bblly=170pt,bburx=340pt,bbury=600pt,width=4.5cm}
\caption{The $\gstar$ cross-section, measured by OPAL,  
compared with model predictions (see text).}
\label{fig6}
\end{center}
\end{figure}

The $\gamma^*\gamma^*$ cross-section has been
advocated as an excellent measurement to see low-$x$ or so called BFKL
effects~\cite{bartels,brodsky}. 
These measurements can be performed by 
tagging  both scattered electrons, which forces the virtuality of
both photons to be in the multi-GeV region. New measurements
from  L3~\cite{L3bfkl} and
OPAL~\cite{OPALbfkl}
of the
double tag cross-section are shown as function of the 'length of the
gluon ladder $Y$'  in Fig.~\ref{fig5} and Fig.~\ref{fig6}.
Here $Y=\ln s/s_0$ and  $s_0= \sqrt{Q_1^2Q_2^2}/(y_1y_2)$.
The measurements are made for the region
$34 <\theta_e <55$  mrad, $E_e> 0.4E_{beam}$  (OPAL)
and 30 mrad $ <\theta_e $, $E_e> 40$ GeV (L3), which leads
to an average $Q^2$ of $\langle 17 \rangle$ (OPAL) and $\langle 15 \rangle$ 
(L3).
L3 observes that the cross-section is larger than the QPM and
PHOJET 1.05 (which does not contain BFKL) predictions.
L3 further fits the $\gamma^*\gamma^*$ cross-section to
$\sigma_{\gamma^*\gamma^*} = \sigma_0/\sqrt{Q_1^2Q_2^2Y}
 \cdot (s/s_0)^{\alpha_{\pom}-1}$, 
 and finds
$\alpha_{\pom} = 0.36\pm 0.02$, i.e.~considerably larger than the pomeron
term in $\gamma\gamma$. The OPAL data are above the QPM and PHOJET1.10
predictions, but much less significantly than the L3 data. 
A 
possible difference
is the sensitivity to radiative corrections of the two 
measurements~\cite{OPALbfkl}. The OPAL measurement was optimized to minimize
these effects.

The LO BFKL calculation  (Bartels99) is
considerably above the data at large $Y$. 
Better  descriptions are found when NLO corrections are taken into
account (Kim99, Kwiecinski et al.)~\cite{kim,kwiecinski}. 
In particular the calculation of ~\cite{kwiecinski},
which predicts moderate BFKL effects in the LEP region, describes the 
data well.

\section{Conclusion}
The total $\gamma\gamma$ cross-section measurements from LEP show an 
intriguing rise with increasing $\Wgg$. There is now some 
evidence that this rise is faster than in hadron-hadron interactions.
In $\gamma^*\gamma^*$ scattering there is evidence in the L3 data 
that the cross-section is above expectation when no BFKL effects are taken into
account. The evidence in the OPAL data is broadly consistent with the 
L3 observation but the significance of the excess over non-BFKL
calculations is  weaker. 

\section{Acknowledgement}
I thank the organizers of this interesting conference for there
hospitality and support, and Stefan
S\"oldner-Rembold for a careful reading of this manuscript.



\begin{thebibliography}{9}
\bibitem{L3sig} 
L3 Collab., M. Acciarri et al.,
     Phys. Lett. {\bf B408} (1997) 450. 
\bibitem{OPALsig} OPAL Collab.,
G. Abbiendi et al, Eur.Phys. J. {\bf C14} (2000) 199.
\bibitem{pythia}
T.~Sj\"ostrand, Comp.~Phys.~Commun.~{\bf 82} (1994) 74; LU-TP-95-20 (1995).
\bibitem{phojet}
R.~Engel and J.~Ranft, Phys.~Rev.~{\bf D54} (1996) 4244.

\bibitem{pancheri}
R.M. Godbole and G. Pancheri, hep-ph/0010104.

\bibitem{L3paper}  L3 Collab., ICHEP 2000, Osaka, P622.
\bibitem{soldnerDIS}
S. S\"oldner-Rembold, hep-ex/0009025.


\bibitem{SAS} S. Schuler and T. Sjostrand, Z. Phys. {\bf C68} (1995) 607.
\bibitem{us1}
A. Corsetti, R.M. Godbole and G. Pancheri, Phys. Lett. {\bf B435} (1998) 441.
\bibitem{bib-DL}
A.~Donnachie and P.~V.~Landshoff, Phys.~Lett. {\bf B296} (1992) 227.
\bibitem{ttwu} T.T. Wu, Mod. Phys. Lett. {\bf A15} (2000) 9.
\bibitem{GLMN} E. Gotsman, E. Levin, U. Maor, E. Naftali, 
Eur. Phys. J. {\bf C14} (2000) 5. 
\bibitem{aspen} M. Block, E. Gregores, F. Halzen and G. Pancheri, 
Phys. Rev. {\bf D60} (1999) 54024.
\bibitem{BKKS}
B. Badelek, M. Krawczyk, J. Kwiecinski and M. Stasto,
hep-ph/0001161. 
\bibitem{bib-pdg}
Review of Particle Physics, Phys.~Rev.~{\bf D54} (1996) 191.
\bibitem{bib-DL2}A. Donnachie, P.V. Landshoff, Phys. Lett. {\bf B437}
(1998) 408.
\bibitem{BFKL} E.A.Kuraev, L.N.Lipatov and V.S.Fadin, { Sov. Phys. JETP}
{\bf 45} (1977) 199; \\
I.I.Balitski and L.N.Lipatov, { Sov. J. Nucl. Phys.}
{\bf 28} (1978) 822.

\bibitem{deroeck} A.M. Cooper, R.C.E. Devenish, A. De Roeck, 
Int. J. Mod. Phys {\bf A13} (1998) 3385.
\bibitem{bartels} J. Bartels, A. De Roeck, H. Lotter
Phys. Lett. {\bf B389} (1997) 742;\\
J. Bartels et al., hep-ph/9710500.
\bibitem{brodsky} S.J. Brodsky, F. Hautmann and D.E. Soper, Phys Rev.
{\bf D56}
(1997) 6957.
\bibitem{L3bfkl} L3 Collab.,  
 M. Acciarri et al., Phys. Lett. {\bf B 453} (1999) 333;
ICHEP 2000, Osaka, P587.
\bibitem{OPALbfkl} M. Przybycien, talk at Photon2000.
\bibitem{kim} S. J. Brodsky et al., JETP Lett. {\bf 70} (1999) 155;
V. Kim, private communication.
\bibitem{kwiecinski} J. Kwiecinski, P. Motyka, hep-ph/0010029.
\end{thebibliography}
\end{document}